\documentclass[a4paper,12pt]{article}

\def\apj{ApJ}                 
\def\apjl{ApJ}                
\def\mnras{MNRAS}             

\usepackage[latin1]{inputenc}
\usepackage[english]{babel}
\usepackage{graphicx}
\usepackage{longtable}
\usepackage{lscape}
\usepackage{natbib}

\begin{document}

\vspace{0.5cm}

\begin{center}
\noindent  THE $V_c \, - \, {\sigma}_0$ RELATION IN LOW SURFACE BRIGHTNESS
GALAXIES: INCLUDING THE LIGHT CONCENTRATION\\

\vspace{0.3cm}

{\large A.Caon, A.Pizzella, {\tiny AND} F.Bertola}\\

{\small Department of Astronomy, University of Padova, Vicolo
dell'Osservatorio 3, I-35122, Padova, Italy}\\
\end{center}

\noindent {\it Poster presented at the international meeting ``Galaxy
Properties across Cosmic Ages'' held in Rome, Accademia dei Lincei,
April 28-29, 2009.}

\vspace{0.5cm}

\noindent
{\Large{\bf Abstract}}\\ 
It has been found that there is a strong correlation
between the circular velocity $V_c$ and the central stellar velocity
dispersion $\sigma_0$ of galaxies. In this respect, low surface
brightness galaxies (LSB) follow a different relation when compared to
Elliptical and high surface brightness (HSB) galaxies. The intrinsic
scatter of the $V_c-\sigma_0$ is partially due to the different
concentration of the light distribution.  In this work we measure the
$C_{28}$ concentration parameter for a sample of 17 LSB finding that
the $C_{28}$ parameter does not account for the different behavior in
the $V_c - {\sigma}_0$ for this class of objects.\\

\noindent
{\Large{\bf Introduction}}\\ 
It has been recently found that the circular velocity $V_c$ measured
in the flat region of the rotation curve and the stellar central
velocity dispersion ${\sigma}_0$ are related
\citep{2002ApJ...578...90F}.  It has successively found that Low
Surface Brightness (LSB) galaxies follow a different $V_c \, - \,
{\sigma}_0$ relation compared with HSB (High Surface Brightness) and E
(Elliptical) galaxies \citep{pizzella:lsb}. It has been suggested
\citep{courteau:c28} that the light concentration of the luminous
component, measured by the $C_{28}$ parameter, may account for this
difference. Aim of this work is to measure the $C_{28}$ for a sample
of LSB galaxies and see if their behaviour in the $V_c \, -
\,{\sigma}_0 - C_{28}$ plane is different with respect to HSB
galaxies.\\

\begin{figure}[h]
\begin{center}
\includegraphics[width=14cm]{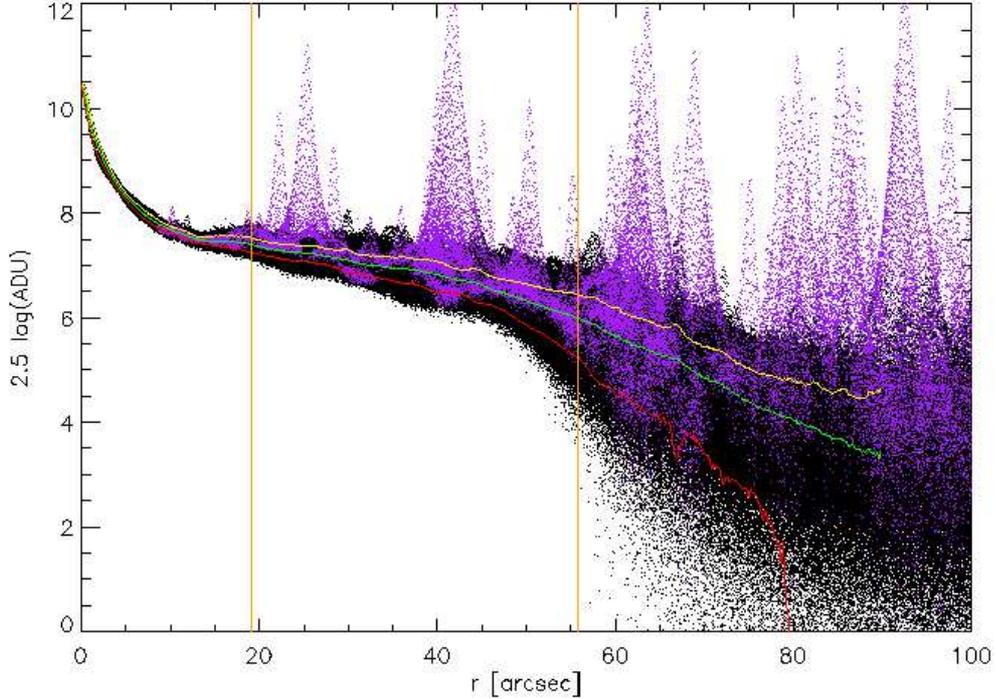}
\end{center}
\caption{Radial surface brightness profile for ESO 514-10. The
  extracted profile is the {\it green line} while the $1\sigma$
  uncertainty are shown as {\it red} and {\it yellow} lines. The
  masked pixels (mostly contaminated by stars) are plotted as {\it
    purple dots}. The radii that contain the $20 \, \%$ and the $80 \,
  \%$ of the total light $r_{20}$ and $r_{80}$ are the {\it orange
    vertical lines}.}
\label{fig:profile_eso_514_10}
\end{figure}

\noindent
{\Large{\bf Data reduction}}\\ 
We selected a sample of 17 LSBs whose circular velocity $V_c$ and
central velocity dispersion of the stellar component ${\sigma}_0$ are
known \citep[][Cardullo {\it Laurea Thesis}
2006]{pizzella:data}. For all the sample galaxies we have broad band
imaging. For each galaxy we derived the radial surface brightness
profile. We first derived, by means of the IRAF task {\tt ellipse},
the photometric center, position angle and ellipticity of the
isophotes. For each pixel of the image we computed its deprojected
distance to the center of the galaxy (that is the distance of the
pixel to the center on the plane of the disk). We derived the mean
surface brightness profile and the $\pm 1\sigma$ uncertainties.  In
this process we automatically masked out the pixel affected by stars.
In Fig. \ref{fig:profile_eso_514_10} you can see the radial surface
brightness profile for ESO 514-10.  We then extrapolated with a
straight line the outer surface brightness profile, computed the total
luminosity and, finally, the radii enclosing the 80\% and 20\% of the
total galaxy light. A Monte Carlo simulation that allows some
variation in the center, position angle and ellipticity of the
isophotes has been used in order to derive the uncertainties in the
$C_{28}$
parameter.
\begin{figure}[h]
\begin{center}
\includegraphics[width=12cm]{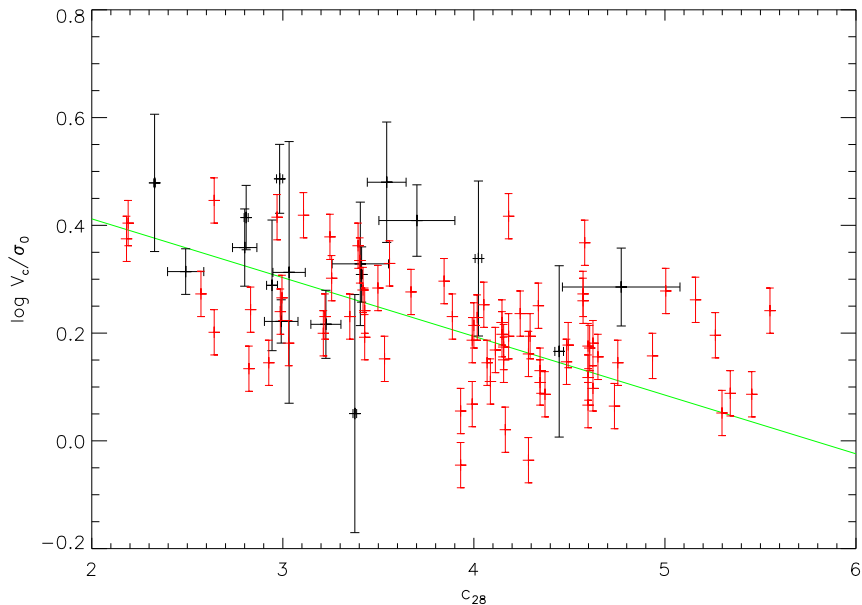}
\end{center}
\caption{$\log V_c / {\sigma}_0$ ratio versus the
concentration parameter $C_{28}$ for our data points ({\it black}) compared
with the data points ({\it red}) and the fit ({\it green line}) found by
\citet{courteau:c28}.}
\label{fig:vs_c28_duo}
\end{figure}

\begin{figure}[h]
\begin{center}
\includegraphics[width=12cm]{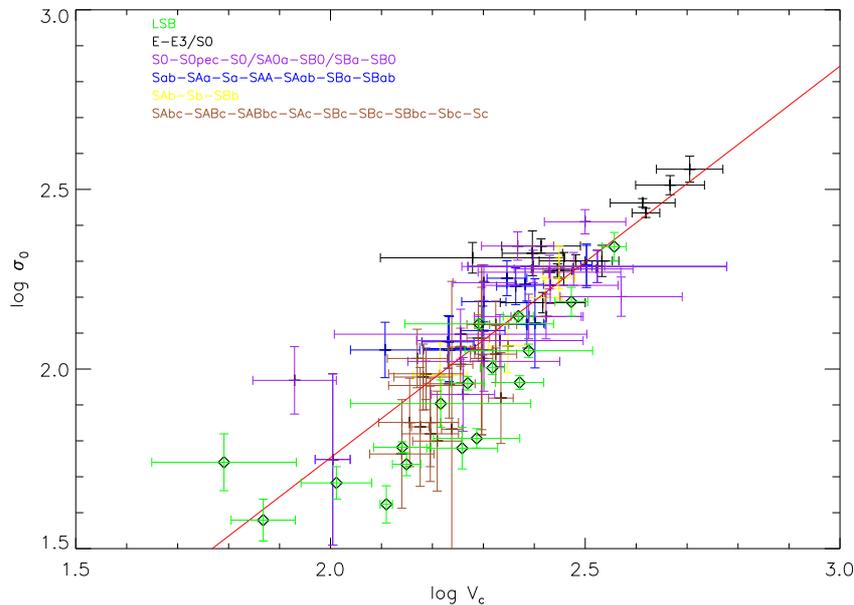}
\end{center}
\caption{$\log V_c$ versus $\log {\sigma}_0$ for the data points of
  the \citet{courteau:c28} sample and for our data points. Different
  colours indicates different morphological types as listed in the
  labels. LSB galaxies ({\it green points}) mostly lies below the
  relation derived by
  \citet{pizzella:lsb}. The {\it red line} indicates the linear
  regression of the HSB data only.}
\label{fig:logs_logv}
\end{figure}

We compared our data points with the data points and the empirical fit
\begin{eqnarray}
\log (V_c / {\sigma}_0) \, = \, 0.63 \, - \, 0.11 C_{28}.
\label{eqn:fit_court}
\end{eqnarray}
found by \citet{courteau:c28} (Fig. \ref{fig:vs_c28_duo}).  The two
distributions seem to follow a similar relation, although the limited
number of points does not allow us to derive an independent 
 linear regression for LSB alone.
\citet{pizzella:lsb} found that in the $\log \, V_c - \log \,
{\sigma}_0$ plane LSB galaxies mostly lies below the relation defined
by the HSB and E galaxies. This is shown in Fig. \ref{fig:logs_logv}
where different morphological types of galaxies are shown.  To better
compare HSB and LSB galaxies, we perform a linear regression of HSB
data points. We used the linear regression described by
\citet{akritas&bershady:lin_regr}. The result of the linear fit is
also shown in Fig. \ref{fig:logs_logv}.\\

\begin{figure}[h]
\begin{center}
\includegraphics[width=12cm]{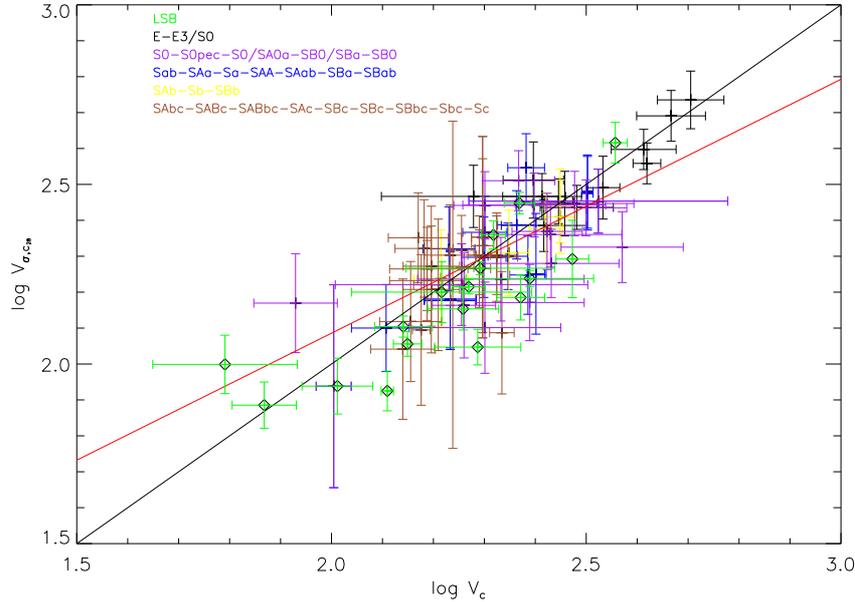}
\end{center}
\caption{$\log V_c$ versus $\log V_{{\sigma},C28}$ derived from
  ${\sigma}_0$ and from $C_{28}$ with the relation empirically found
  by \citet{courteau:c28}. The {\it red line} indicates the linear
  regression of the HSB data points while the {\it black line} is the $V_c
  = V_{{\sigma},C28}$ line.}
\label{fig:v_v}
\end{figure}

\noindent
{\Large{\bf Results}}\\ 
Now that we have measured the $C_{28}$ parameter for LSB galaxies we
used the empirical relation found by \citet{courteau:c28} and shown in
Fig. \ref{fig:vs_c28_duo} to derive their $V_{{\sigma},C28}$ from
${\sigma}_0$ and $C_{28}$.  In Fig. \ref{fig:v_v} we plot our new LSB
points together with the points derived by \citet{courteau:c28} for
HSB galaxies. Applying a linear regression we found
\begin{eqnarray}
\log \, V_{{\sigma},C28} \, = \, 0.672 + 0.707 \, \log \, V_c
\label{eqn:fit_v_v}.
\end{eqnarray}
LSB galaxies still seems follow a different relation, having, for a given $V_c$
a smaller value of $V_{{\sigma},C28}$.

\begin{figure}[h]
\begin{center}
\includegraphics[width=67mm]{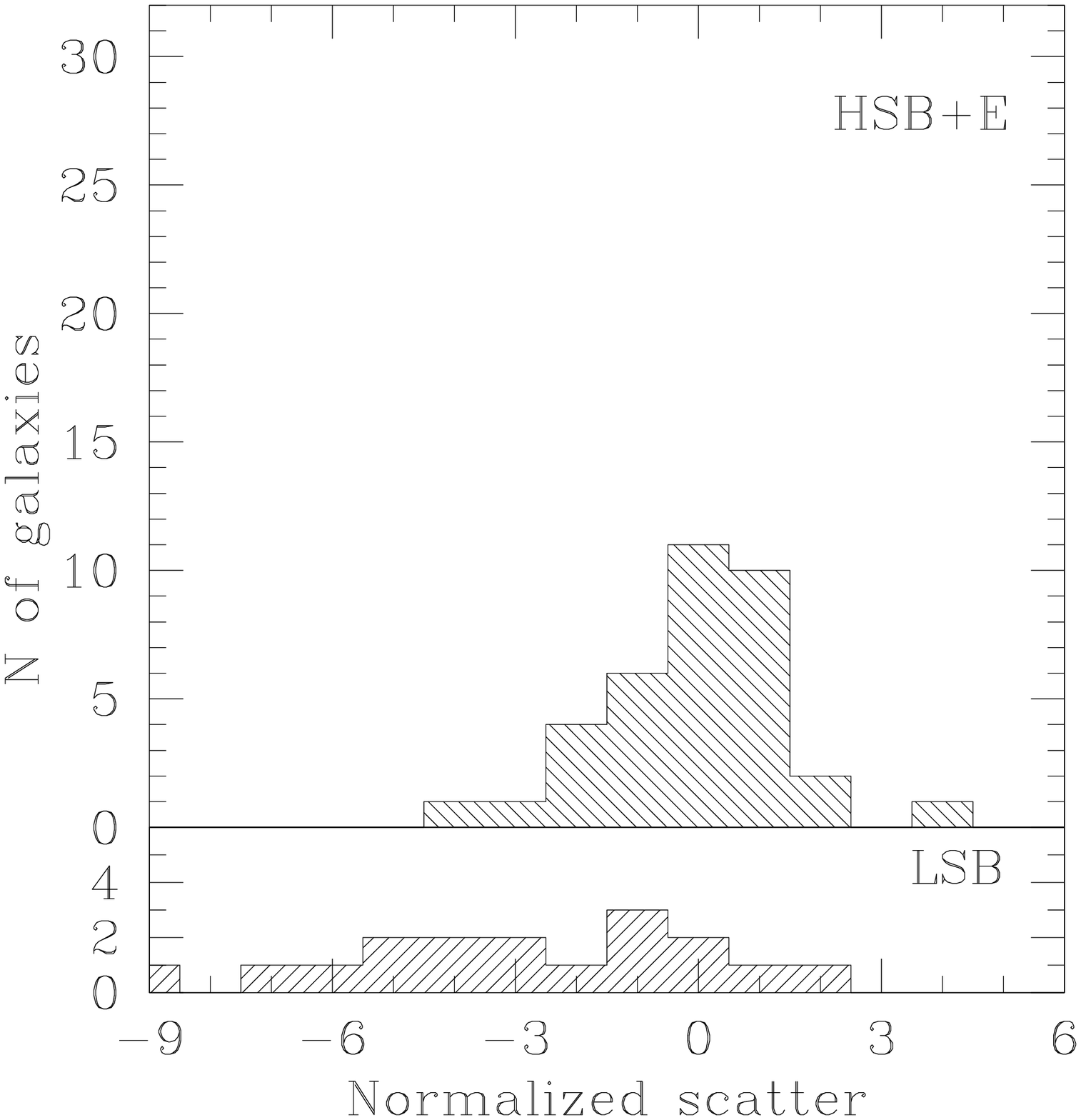}
\includegraphics[width=67mm]{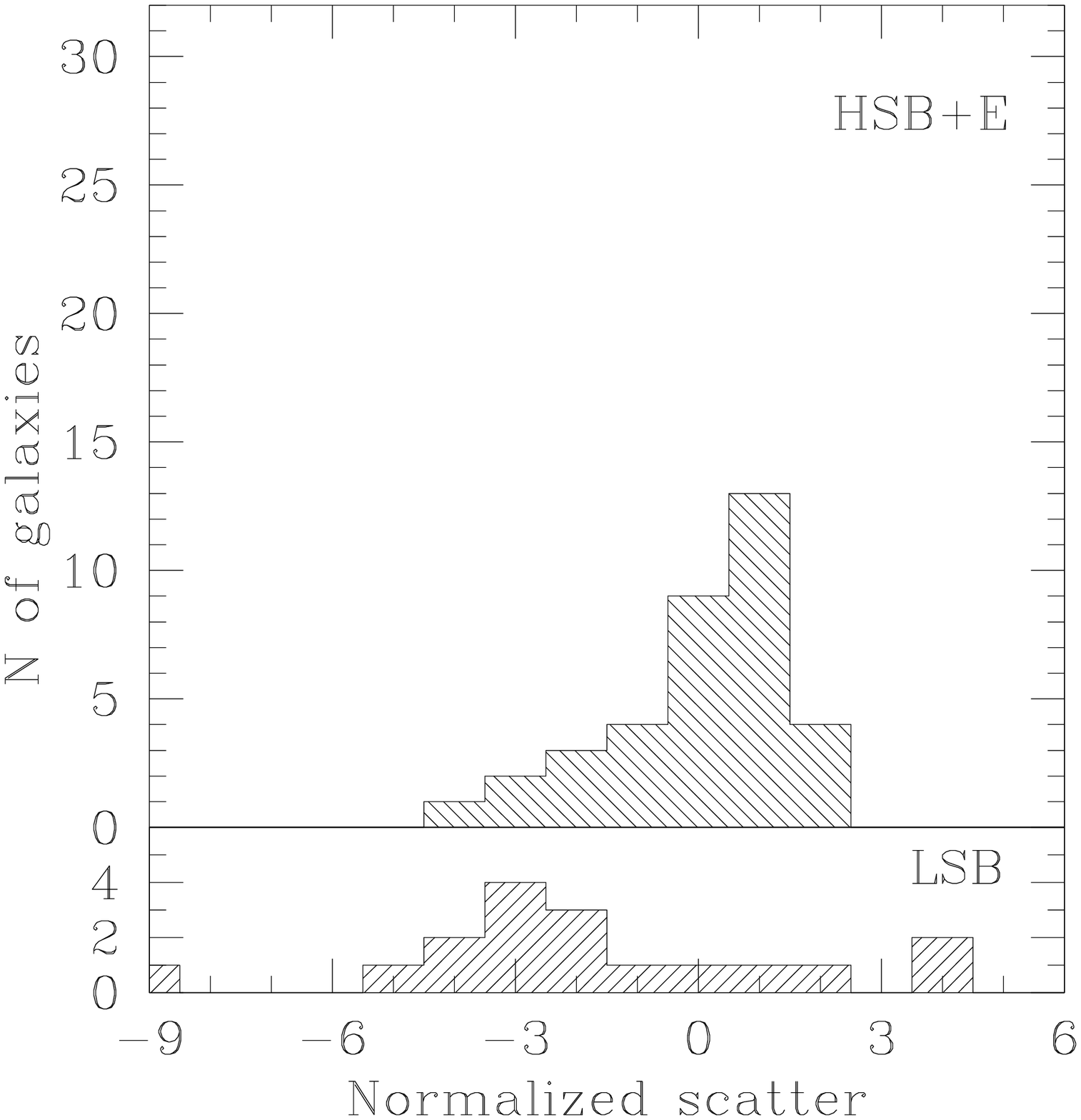}
\end{center}
\caption{{\it Left panel}: Histograms of the scatter from the linear
 regression for HSB and LSB in the $ V_c - {\sigma}_0$ plane
 (Fig. \ref{fig:logs_logv}). We can note that our LSB data points
 mostly lies only in the negative side of the distribution indicating that
 these galaxies follow a different relation. {\it Right panel}: As
 left panel in the $ V_c - V_{{\sigma},C28}$ plane.  Again, LSB show a
 different distribution.}
\label{fig:scarti_logs_logv}
\end{figure}

Let us remind that our aim is to test whether LSB and HSB share the
same $V_c \, - \, {\sigma}_0 - C_{28}$ relation. The best way of doing
it is to analyze the distribution of the scatter of the HSB and LSB
measurements with respect to the linear regression defined by the HSB
sample. For each data point we computed the scatter as done by
\citet{pizzella:lsb}, that is computing the difference between the
observed value and the linear regression and normalizing it by means
of its observational uncertainties. The scatter distribution of HSB
and LSB is then compared by means of a KS test.  We performed this
analysis both on the $V_c \, - \, {\sigma}_0$ and $V_c \, - \,
{\sigma}_0 - C_{28}$ relations in order to undestand how the two
relations compare.  The plot of the residuals is shown in
Fig.\ref{fig:scarti_logs_logv}. 
The result of the KS test is that LSB and HSB have two distinct
distributions both considering the $V_c \, - \, {\sigma}_0$ relation
(and this is basically the result obtained by \citet{pizzella:lsb}
with a similar data set) and considering the $V_c \, - \, {\sigma}_0 -
C_{28}$ relation.\\

\noindent
{\Large{\bf Conclusions}}\\ 
The KS test indicates that the distribution of HSB and LSB are
different at a $99.9$\% and $99.8$\% confidence level in the $ V_c -
{\sigma}_0$ and $ V_c - V_{{\sigma},C28}$ plane respectively.  Our
conclusion is therefore that, concerning LSB galaxies, the $C_{28}$ parameter
does not account for the different behaviour in the $V_c - {\sigma}_0$
relation. Since ${\sigma}_0$ is related to the SMBH mass and $V_c$ is
related to the dark matter halo mass, this result may indicate that
either LSB galaxies follow a different $M_{SMBH} - {\sigma}_0$
relation with respect to HSB galaxies, or alternatively, that LSB
galaxies have SMBH of smaller mass compared with HSB galaxies.


\end{document}